\documentstyle[preprint,aps]{revtex}
\begin{document}
\draft

\preprint{\parbox{2in}{
\begin{flushright} UMHEP-418 \\ hep-ph/9505256 \end{flushright}
}}

\title{Observing the technieta at a photon linear collider}
\author{ Jusak Tandean }
\address{Department of Physics and Astronomy, University of Massachusetts,
Amherst, Massachusetts 01003}

%\date{\today}
\maketitle

\begin{abstract}
If electroweak symmetry is spontaneously broken by technicolor, there will be
technimesons analogous to the mesons of the ordinary strong interactions.
One of the lightest technimesons is the technieta $\eta'_T$ (the analogue of
the~$\eta'$).
In this work we consider the possibility of observing it at a linear $e^+e^-$
collider operating in the $\gamma\gamma$ mode.
Detecting the process $\gamma\gamma\rightarrow\eta'_T$ allows for
the measurement of the $\eta'_T\gamma\gamma$ anomalous coupling which is in
principle sensitive to the underlying technifermion substructure.
\end{abstract}


%\narrowtext
%\widetext

The mechanism which is responsible for the spontaneous breaking of
$SU(2)_L\otimes U(1)_Y$ electroweak symmetry down to the $U(1)$ of
electromagnetism remains unknown.
In the standard electroweak model this mechanism involves elementary scalar
fields, whose presence in the theory makes it in some sense `unnatural'
\cite{ls}.
One of the proposed alternatives to the standard model is technicolor
\cite{efls}.
In this scenario, there are no elementary scalar fields, and electroweak
symmetry is spontaneously broken by condensates of new fermions bound by the
technicolor forces.

In the minimal technicolor (TC) model \cite{efls}, in addition to the usual
quarks and leptons, there are two new fermions $\cal U$ and $\cal D$.
These technifermions engage in TC and electroweak interactions, but are
color singlets.
The TC dynamics is assumed to be QCD-like and have an $SU(N_{\rm TC})$ gauge
symmetry.
Each of the technifermions is assigned to the fundamental
(\(\mbox{\boldmath $N$}_{\rm TC})\) representation of $SU(N_{\rm TC}).$
With respect to the electroweak $SU(2)_L\otimes U(1)_Y$ gauge group, the
left-handed components of the technifermions form a weak-isospin doublet
$\,{\cal Q}_L = ({\cal U}_L,{\cal D}_L)\,$ having zero hypercharge, whereas the
right-handed components ${\cal U}_R$ and ${\cal D}_R$ are weak-isospin singlets
with hypercharge  $\;Y_R=1~{\rm and}~-1\,$, respectively.
It follows that the charges of $\cal U$ and $\cal D$ are
$\;Q_{\cal U} = {1\over2}\;$ and $\;Q_{\cal D} = -{1\over2}\;.\;$
With these assignments, the model is free of gauge anomalies.

In the absence of electroweak interactions, the TC Lagrangian in the massless
limit has a global techniflavor $SU(2)_L \otimes SU(2)_R$ symmetry.
The situation is analogous to the chiral limit of QCD.
When the TC forces become strong at an energy scale
$\;\Lambda_{\rm TC}={\cal O}(1\rm\,TeV)\;,\;$ they form technifermion
condensates which break the global symmetry down to $SU(2)_{L+R}$.
As a consequence, there are three Goldstone bosons, which are the massless
technipions $\pi_T^{\pm,0}$, corresponding to the broken generators.
The axial currents corresponding to these generators couple to the
$\pi_T^{\pm,0}$ with strength $\;F_{\pi_T}=246\,\rm GeV\;.\;$
When the electroweak interactions are turned on, the electroweak gauge bosons
couple to the axial currents.
Consequently, the $W^\pm$ and $Z^0$ gauge bosons acquire masses, and the
$\pi_T^{\pm,0}$  disappear from the physical spectrum, having become the
longitudinal components of the $W^\pm$ and $Z^0$.

Since the minimal TC model cannot account for the masses of ordinary quarks and
leptons, generating their masses requires new interactions.
Hence, in considering the minimal TC model in this paper, we assume that the
model is a low energy sector of a larger, more realistic model incorporating
such interactions, which occur at scales  higher than several TeV \cite{rs}.
Since here we are working at energies up to only a few TeV, we expect that we
can ignore these interactions.
At such energies these interactions are assumed to be represented by effective
four-Fermi interactions involving technifermions and ordinary fermions, so that
when technifermion condensates form, ordinary fermions get their masses.

The lightest technihadrons in the physical spectrum can be inferred from QCD.
The $\pi_T^{\pm,0}$ having been absorbed by the $W^\pm$ and $Z^0$, the spectrum
begins with a technieta $\eta'_T$, a technirho $\rho_T$, and a techniomega
$\omega_T$.
The $\eta'_T$ and the $\omega_T$ are techni-isospin singlets, while the
$\rho_T$
is a techni-isospin triplet.
They are roughly $\Lambda_{\rm TC}/\Lambda_{\rm QCD}$ times  heavier than their
QCD counterparts.
Information about the substructure of these technihadrons may be obtained by
observing them at multi-TeV colliders.
The production and detection of the $\rho_T$ and the $\omega_T$ have been much
dicussed in the literature \cite{rho}.

In this paper we are concerned with the $\eta'_T$, which has not received much
attention recently.
One detailed study on the $\eta'_T$ of which we are aware was done
more than a decade ago \cite{pdgv}\footnote{One recent paper \cite{cg} dealt
briefly with  the $\eta'_T$ and the possibility of observing it as one of the
decay products of the $\rho_T$ in TC models with scalars}.
Our purpose here is to take another look at the $\eta'_T$ and its detection in
the light of some recent development in collider physics.
One of its decay modes is  $\eta'_T\rightarrow\gamma\gamma$  (in analogy to
$\eta,\pi^0 \rightarrow\gamma\gamma$) which occurs via the anomaly \cite{pdgv},
with technifermions appearing in the loop.
This implies that the observation of the $\eta'_T$ at a $\gamma\gamma$ collider
can be a means to probe the technifermion substructure, providing information
complementary to that found in  $\rho_T$ and $\omega_T$ studies.
Here we shall consider the detection of the $\eta'_T$ at a future high-energy
photon linear collider realized by laser-backscattering technique at a linear
$e^+e^-$ collider \cite{ifg}.
Such a $\gamma\gamma$ collider produces high-luminosity beams and clean
backgounds, which are necessary factors in this endeavor because of the
smallness of the   $\eta'_T\gamma\gamma$ coupling.

Our ability to detect $\gamma\gamma\rightarrow\eta'_T\rightarrow X$ depends on
the choice of final state $X$.
The four-Fermi couplings between technifermions and ordinary fermions may cause
the $\eta'_T$ to have appreciable decays into a pair of ordinary fermions.
Since a reliable model for the couplings of  the  $\eta'_T$ to ordinary
fermions
is still unknown, conventional wisdom, based on analogy with the minimal
standard model, suggests that  they are proportional  to fermion mass, and so
the $\eta'_T$ will decay mostly to a pair of top quarks.
We shall show that in such a case the $t\bar{t}$ decay mode can provide a
window
to observe the $\eta'_T$.
Now, we have no guarantee that the tendency to couple to fermion mass will
always occur; for even in multiple-Higgs models such a tendency can be
evaded.
Hence we shall also consider the possibility that the decays into fermion pairs
are negligible.
Consequently, the $\eta'_T$ decays mostly into two and three electroweak
gauge bosons and becomes a very narrow resonance.
In this case we shall show that the $\eta'_T$ is detectable in its
$\gamma\gamma$ decay mode.

In order to discuss the decays of the  $\eta'_T$, we need to construct the
relevant effective lagrangians.
The effective lagrangian which gives the couplings of the $\eta'_T$ to
ordinary fermion pairs can be written as
\begin{eqnarray}
{\cal L}_{\eta'_T f\bar{f}}&=&
 -i \sum_f \lambda_f {m_f\over F_{\pi_T}} \;
  \bar{\psi}_f \, \gamma_5 \, \psi_f \, \eta'_T \;,
\end{eqnarray}
where $\psi_f$ and $m_f$ are, respectively, the fields and masses of the
fermions, $\lambda_f$ are  dimensionless constants, and the sum is over all
ordinary leptons and quarks.
For the case in which the $\eta'_T$ couples  to fermion mass, we shall take
$\;\lambda_f = 1\;$  for all $f$'s,  resulting in the same  lagrangian as that
in Ref.~\cite{pdgv}.
In the second case, in which the decays into fermion pairs are negligible,  we
shall set $\;\lambda_f = 0\;$  for all $f$'s.

In the second case the decays of $\eta'_T$ into electroweak gauge bosons become
important.
Since the minimal TC dynamics is QCD-like, we expect that some of the
decay  properties of the $\eta'_T$ are similar to those of the $\eta$ and
$\eta'$ of  QCD.
They are two of the lightest pseudoscalar mesons whose low-energy interactions
with the lightest vector mesons and the photon can be described by an effective
chiral Lagrangian \cite{tf,js,ugm}.
What we need then is an analogous effective Lagrangian which can describe
interactions involving the lightest pseudoscalar and vector technimesons as
well
as the electroweak gauge bosons at energies below $\Lambda_{\rm TC}$.

Before writing down the desired effective Lagrangian \cite{js'}, we discuss the
fields contained in it.
We collect the pseudoscalar fields $\eta'_T$ and \(\mbox{$\boldmath\pi_T$}\)
into a unitary matrix $\;U=\exp(i\varphi/F_{\pi_T})\;,\;$          where
\(\;\varphi = 1\eta'_T + \mbox{\boldmath $\tau \cdot \pi_T$}\;,\;\)  with
$\;\mbox{\boldmath $\tau$}=(\tau^1,\tau^2,\tau^3)\;$ being Pauli matrices.
This construction has been chosen because an analogous choice made in QCD leads
to good agreement with data, as will be noted later.
Under global $U(2)_L\otimes U(2)_R$ transformations,
$\;U\rightarrow LUR^\dagger\;$ with $\;L,R \in U(2)\;.$
For the vector technimesons
\(\omega_T^\mu\) and  \(\mbox{\boldmath$\rho_T^\mu$}\), we construct a matrix
\(\;V^\mu = {1\over2} g
 (1\omega_T^\mu + \mbox{\boldmath $\tau \cdot \rho_T^\mu$})\;,\;\)
where $g$ is a constant.
The relation between $g$ and the $\rho_T\pi_T\pi_T$ coupling constant will be
shown below.
$V_\mu$ is related to auxiliary fields $A^{L,R}_\mu$ defined by
$\;A^L_\mu = \xi V_\mu \xi^\dagger + i\partial_\mu \xi\,\xi^\dagger \;$ and
$\;A^R_\mu = \xi^\dagger V_\mu \xi + i\partial_\mu \xi^\dagger \xi \;,\;$
where $\;\xi = U^{1/2} \;.\;$
In order to include the electroweak interactions, we gauge an
$SU(2)\otimes U(1)$ subgroup of the global $U(2)_L \otimes U(2)_R$, and
introduce left-handed and right-handed gauge fields
\(\;\ell_\mu={1\over2}g_2\mbox{\boldmath $\tau \cdot W_\mu$}\;\) and
 $\;r_\mu={1\over2}g_1\tau^3 B_\mu \;,\;$ reflecting the assignments
$\;Q_{\cal U} = {1\over2}\;$ and  $\;Q_{\cal D} = -{1\over2}\;$ of the
underlying technifermions.
$B_\mu$ and \(\mbox{\boldmath $W_\mu$}\) are defined in
terms of the photon, $W^\pm$, and $Z^0$ fields in the standard way.
Under $SU(2)_L\otimes U(1)_Y$ gauge transformations,
\begin{equation}
\ell_\mu\rightarrow L\ell_\mu L^\dagger + i\partial_\mu L\,L^\dagger \;,\;\;\;
r_\mu\rightarrow Rr_\mu R^\dagger + i\partial_\mu R\,R^\dagger \;,
\end{equation}
where now $\;L\in SU(2)_L\;$ and $\;R = \exp(-{1\over2}i\tau^3\theta) \;.\;$
We require $A^L_\mu\,(A^R_\mu)$ to transform the same way as
$\ell_\mu\,(r_\mu),$
\begin{equation}
A^L_\mu\rightarrow LA^L_\mu L^\dagger + i\partial_\mu L\,L^\dagger \;,\;\;\;
A^R_\mu\rightarrow RA^R_\mu R^\dagger + i\partial_\mu R\,R^\dagger \;.
\end{equation}
This is to ensure gauge invariance, as we will see shortly.

We divide into four parts the effective Lagrangian $\cal L$ which gives the
decays of the  $\eta'_T$ into electroweak gauge bosons.
The corresponding action is written as
\begin{equation}
\Gamma = \int{\cal L}\,d^4x
       = \int ({\cal L}_1+{\cal L}_2)\,d^4x + \Gamma_3 +\Gamma_4 \;.
\end{equation}
We have included in $\cal L$ only the most relevant terms for our purposes.
To $\cal L$ must be added the kinetic terms of the vector technimesons and the
electroweak gauge bosons, which we will not write explicitly.

The first part of $\cal L$ is\footnote{What we have here is analogous to the
approximate vector-meson dominance mentioned in Ref.~\cite{js}.}
\begin{eqnarray}
{\cal L}_1&=&
 {m^2\over 2g^2}\,{\rm Tr}\Bigl[(A^{L\mu}-\ell^\mu)(A^L_\mu-\ell_\mu)
                            +(A^{R\mu}-r^\mu)(A^R_\mu-r_\mu)\Bigr] \nonumber\\
 &&-\;{b\over 2g^2}\,{\rm Tr}
       \Bigl[(A^{L\mu}-\ell^\mu)U(A^R_\mu-r_\mu)U^\dagger\Bigr] \;,
\end{eqnarray}
which is evidently gauge invariant.
Upon expanding to second order of pseudoscalar fields, from the vector
mass term and the pseudoscalar kinetic term in ${\cal L}_1$ we obtain the
relations
\begin{eqnarray}
m_{\rho_T}^2 = m_{\omega_T}^2 \;,\;\;\;
\;\;m^2 = {1\over2}(m_{\rho_T}^2+g^2 F_{\pi_T}^2) \;,\;\;\;
b = g^2 F_{\pi_T}^2 - m_{\rho_T}^2 \;.
\label{m}
\end{eqnarray}
We could also obtain in ${\cal L}_1$ the mass terms of the $W^\pm$ and $Z^0$.
Using (\ref{m}), we find in ${\cal L}_1$ two terms which will be pertinent in
evaluating the decays of the $\eta'_T$:
\begin{equation}
{\cal L}_{\rho_T\pi_T\pi_T} = {i\over2}\,{g_{\rho_T\pi_T\pi_T}\over g}\,
{\rm Tr} \Bigl[V^\mu(\varphi \partial_\mu\varphi
                     -\partial_\mu\varphi\,\varphi)\Bigr] \;,
\label{vpp}
\end{equation}
where\footnote{Setting  $\; g = g_{\rho_T\pi_T\pi_T} \;$  would give us the TC
counterpart of the KSRF \cite{ksrf} relation.}
$\;g_{\rho_T\pi_T\pi_T} = m_{\rho_T}^2/(2g F_{\pi_T}^2) \;,\;$  and
\begin{equation}
{\cal L}_{\rho_T{\cal A}} =
-{m_{\rho_T}^2\over g^2}\,{\rm Tr} \Bigl[V^\mu(\ell_\mu+r_\mu)\Bigr] \;,
\label{va}
\end{equation}
which gives the couplings of the vector technimesons to the electroweak gauge
bosons, $\;{\cal A}=(\gamma,W^\pm,Z^0) \;.\;$
The mass of the $\rho_T$ and $g_{\rho_T\pi_T\pi_T}$ can be expressed in terms
of their QCD counterparts by using large-$N_{\rm TC}$ arguments and scaling
from
QCD \cite{sd}.
One gets
\begin{eqnarray}
m_{\rho_T} =
{F_{\pi_T}\over F_\pi}\Biggl({3\over N_{\rm TC}}\Biggr)^{1\over 2} m_\rho
\;,\;\;\;
g_{\rho_T\pi_T\pi_T} =
 \Biggl( {3\over N_{\rm TC}} \Biggr)^{1\over2}\,g_{\rho\pi\pi} \;,
\label{grpp}
\end{eqnarray}
where $\;F_\pi = 92\,\rm MeV\;$ is the pion decay constant,
$\;m_\rho = 770\,{\rm MeV}\;$ is the $\rho$-meson mass,  and
$g_{\rho\pi\pi}$ is related to the decay  width of the $\rho$  by
$\;\Gamma_{\rho\rightarrow\pi\pi}=
 g_{\rho\pi\pi}^2\,|{\rm\bf p}_\pi|^3/ (6\pi m_\rho^2) \;.\;$

The second Lagrangian expresses the breaking of the axial $U(1)$ symmetry and
provides mass for the $\eta'_T$ in the chiral limit.
It is written as
\begin{equation}
{\cal L}_2 =  {a F^2_{\pi_T}\over8 N_{\rm TC}}\,
 \Bigl[{\rm Tr}({\rm ln}U-{\rm ln}U^\dagger)\Bigr]^2 \;,
\label{l2}
\end{equation}
with $a$ being a constant.
$a$ is connected to the mass $m_{\eta_0}$ of the flavor $SU(3)$ singlet of QCD
in the chiral limit by \cite{pdgv}
\begin{equation}
a = {1\over6}{F^2_{\pi_T}\over F^2_\pi}{9\over N_{\rm TC}}\,m_{\eta_0}^2 \;,
\label{a}
\end{equation}
where  \cite{dgh}  $\;m_{\eta_0} \simeq 849\,\rm MeV\;$.
Assuming massless technifermions, we find from (\ref{l2}) and (\ref{a}) that
the
mass of the $\eta'_T$ is
\begin{equation}
m_{\eta'_T} =
\sqrt{{2\over3}}{F_{\pi_T}\over F_\pi}{3\over N_{\rm TC}}\,m_{\eta_0}
\simeq {4\over N_{\rm TC}} \,1.39\,\rm TeV \;.
\end{equation}
Hence the $\eta'_T$ may be the lightest technihadron in the physical spectrum.

The third and fourth parts in $\Gamma$ contain terms proportional to the
Levi-Civita tensor, $\epsilon_{\kappa\lambda\mu\nu}$.
We find it convenient to write them compactly using the notation of
differential forms \cite{js}.
Hence we define  $\;dU = \partial_\mu U \,dx^\mu \;,$
$\;\alpha = dU\,U^\dagger \;,$   $\;\beta = U^\dagger\alpha U \;;\;$ and
$\;{\cal V} = {\cal V}_\mu \,dx^\mu\;$ and
$\;d{\cal V}=\partial_\mu {\cal V}_\nu \, dx^{\mu} dx^{\nu}\;$
for $\;{\cal V}=A^L,\ell,r \;.\;$
The third piece, $\Gamma_3$, is the gauged Wess-Zumino-Witten action
\cite{wzw}.
Written in terms of differential forms, it is
\begin{eqnarray}
\Gamma_3 &=& \Gamma_{WZW}(U,\ell,r)   \nonumber\\
&=& c\int{\rm Tr}\bigl[
          (d\ell \,\ell+\ell \,d\ell)\alpha + (dr\,r+r\,dr)\beta
          - d\ell \,dU\,rU^\dagger + dr\,dU^\dagger\,\ell U\bigr] \nonumber\\
&&+\;ci\int{\rm Tr}(\ell \alpha^3 + r \beta^3) + \cdots
\label{g3}
\end{eqnarray}
where $\;c=-iN_{\rm TC}/(48 \pi^2)\;$ and we have shown only the relevant
terms.
For our choice of quantum numbers of the underlying technifermions, $\Gamma_3$
is
gauge invariant.
In order to write $\Gamma_4$, we need additional differential forms.
They are
$\;\alpha_1 = \alpha + iA^L - iUrU^\dagger \;,\;$  $\;\alpha_2 = -iA^L +
i\ell \;,\;$   $\;\beta_{1,2} = U^\dagger\alpha_{1,2} U \;,\;$  and the
field-strength two-forms  $\;F({\cal V}) = d{\cal V} + i{\cal V}{\cal V}\;$
for $\;{\cal V}=A^L,\ell,r \;.\;$
It is straightforward to show that
$\;\alpha_{1,2}\rightarrow L\alpha_{1,2}L^\dagger\;$ and
$\;\beta_{1,2}\rightarrow R\beta_{1,2}R^\dagger\;$  under gauge
transformations.
The fourth part is then
\begin{eqnarray}
\Gamma_4 &=& \int{\rm Tr}\Biggl[
 {c_1\over g}(\alpha^3_1 \alpha_2 - \alpha^3_2 \alpha_1)
  + {ic_2\over g^2}F(A^L)[\alpha_1,\alpha_2]
  + \Bigl({c_1\over g}+{c_2\over g^2}-{c_3\over g^3}\Bigr)
     \alpha_1 \alpha_2 \alpha_1 \alpha_2  \nonumber\\
&&\;\;\;\;\;\;\;\;\;
  +\;d_1\Bigl(F(\ell)[\alpha_1,\alpha_2]+F(r)[\beta_1,\beta_2]\Bigr)
\Biggr] \;,
\label{g4}
\end{eqnarray}
where $c_1$, $c_2$, $c_3$, and $d_1$ are constants whose values will be
discussed
below.
We easily see that each of the $c_1$, $c_2$, $c_3$, and $d_1$ terms is gauge
invariant.

The partial decay widths of the $\eta'_T$ can now be evaluated.
In the case of $\eta'_T$ coupling to ordinary-fermion pairs, we set
$\;\lambda_f = 1\;$ for all $f$'s in ${\cal L}_{\eta'_T f\bar{f}}$, from which
amplitudes for the decays into fermion pairs can be easily extracted.
In this case the $\eta'_T$ also decays into a pair of gluons through a quark
triangle-loop, and the decay amplitude is \cite{etagg}
\begin{eqnarray}
{\cal M}_{\eta'_T \rightarrow g_a g_b} =
{\alpha_s \over \pi F_{\pi_T}}
\sum_q {-1 \over 2 R_q^2}
 \bigg[ 2\ln{\Big( {\mbox{$1\over2$}}R_q+\sqrt{{\mbox{$1\over4$}}
R^2_q-1}\Big)}
       + i\pi \bigg]^2
\delta_{ab} \;\epsilon_{\kappa\lambda\mu\nu}\,
 k_1^\kappa k_2^\lambda \,\epsilon^{\mu\ast}(k_1) \epsilon^{\nu\ast}(k_2)  \;,
\end{eqnarray}
where  $\;a,b = 1,...,8\;$  are gluon color indices, the sum is over all
quarks, $\;R_q = m_{\eta'_T}/m_q\;,\;$  and  the strong coupling constant
$\;\alpha_s = \alpha_s (m_{\eta'_T}) \;.\;$
In order to compute the  widths of the $f\bar{f}$ and $gg$ decay modes, we
use the values of fermion masses\footnote{We use central mass values in the
case of  quarks.}
available in Ref.~\cite{rpp}  and take  $\;m_t = 174\,\rm GeV\;.\;$
In Table \ref{widths} we summarize the results,
along with those of the decays into  electroweak gauge bosons to be discussed
below, for  $\;N_{\rm TC} = 4,5,6\;.\;$
The contribution of the $gg$ mode to the total width is seen to be comparable
to  that of   the decays into electroweak bosons.
It is worth mentioning  that in TC models with  technifermions carrying
ordinary color  we would expect the $gg$ decay mode to proceed mainly through a
colored-technifermion loop, resulting in a much larger $gg$ partial width,
although the $t\bar{t}$ mode would still be dominant  \cite{ctc}.

The decays into two and three electroweak gauge bosons come from
(\ref{g3}) and (\ref{g4}), together with   (\ref{vpp}) and (\ref{va}).
We express the amplitudes for the decays into two (transverse) electroweak
gauge bosons as
\begin{equation}
{\cal M}_{\eta'_T \rightarrow {\cal A}_1{\cal A}_2} =
-C_{\eta'_T \rightarrow {\cal A}_1{\cal A}_2}\,
{e^2 N_{\rm TC}\over8\pi^2 F_{\pi_T}}\, \epsilon_{\kappa\lambda\mu\nu}\,
 k_1^\kappa k_2^\lambda \,\epsilon^{\mu\ast}(k_1) \epsilon^{\nu\ast}(k_2)
\label{egg}
\end{equation}
where
$\;{\cal A}_1{\cal A}_2=W^+ W^-,\,\gamma\gamma,\,\gamma Z^0,\,Z^0 Z^0 \;,\;$
$\;\epsilon_{0123} = +1\;,\;$  and
\begin{eqnarray}
C_{\eta'_T \rightarrow W^+ W^-}={1\over 3s_W^2} \;, &\;\;
C_{\eta'_T \rightarrow \gamma\gamma}=1 \;, &\;\;
C_{\eta'_T \rightarrow \gamma Z^0}={1-2s_W^2\over 2c_W s_W} \;, \nonumber\\&
C_{\eta'_T \rightarrow Z^0 Z^0}={1-3s_W^2+3s_W^4\over 3c_W^2 s_W^2} \;,
\nonumber
\end{eqnarray}
with $\;s_W^2 = \sin^2\theta_W\;$ ($\theta_W$ is the Weinberg angle) and
$\;c_W = \sqrt{1-s_W^2} \;.\;$
The $\eta'_T$ also decays into two technipions and one (transverse) electroweak
gauge boson, in analogy to $\;\eta\rightarrow \pi^+\pi^-\gamma\;$ in QCD.
Since the $\pi_T^{\pm,0}$ have become the longitudinal components of
the  $W^\pm$ and $Z^0$, the amplitudes for these decays are equivalent to those
of the  decays into two longitudinal and one transverse electroweak gauge
bosons,
up to  corrections which vanish at energies much higher than $m_{W,Z}$
\cite{et}.
The amplitudes involving the $\pi_T^{\pm,0}$ are  written as
\begin{equation}
{\cal M}_{\eta'_T \rightarrow \pi_T \pi'_T {\cal A}} =
-C_{\eta'_T \rightarrow \pi_T\pi'_T{\cal A}}\, M(p,p')\,
\epsilon_{\kappa\lambda\mu\nu}\,
 p^\kappa p'^\lambda k_{\cal A}^\mu \,\epsilon^{\nu\ast}(k_{\cal A})
\label{eppg}
\end{equation}
with
$$
M(p,p') = {e\over F_{\pi_T}^3}\Biggl[
{N_{\rm TC}\over 12\pi^2}
 +\Biggl( {ic_2\over g^2}+2d_1 \Biggr) \;
   {(p+p')^2\over (p+p')^2-m_{\rho_T}^2} \Biggr] \;,
$$
where $\;\pi_T\pi'_T{\cal A}=\pi_T^+\pi_T^-\gamma,\,\pi_T^\mp \pi_T^0 W^\pm,\,
\pi_T^+\pi_T^- Z^0,\;$   and
\begin{eqnarray}
C_{\eta'_T \rightarrow \pi_T^+ \pi_T^- \gamma} = 1 \;,\;\;\;
C_{\eta'_T \rightarrow \pi_T^\mp \pi_T^0 W^\pm} = \pm\,{1\over 2s_W} \;,\;\;\;
C_{\eta'_T \rightarrow \pi_T^+ \pi_T^- Z^0} = {1-2s_W^2\over 2c_W s_W} \;,
\nonumber
\end{eqnarray}
with each amplitude coming from direct and $\rho_T$-mediated diagrams, and the
$\pi_T^\pm$ and $\pi_T^0$ now being regarded as longitudinal $W^\pm$ and $Z^0$.
To determine $c_2$ and $d_1$, we make the choice that
\begin{equation}
{ic_2\over g^2} = 2d_1 = -{N_{\rm TC}\over 16\pi^2} \;,
\label{c2}
\end{equation}
for we have found that a similar choice made for the analogous effective
Lagrangian in QCD resulted in good agreement with data for
$\;\eta\rightarrow\pi^+\pi^-\gamma \;,\;$
$\;\rho^\pm\rightarrow\pi^\pm\gamma \;,\;$
and  \cite{tf}   $\;\omega \rightarrow\pi^0\gamma\;$.

The decay widths calculated from (\ref{egg}) and (\ref{eppg})  are listed in
Table \ref{widths}  for different values of $N_{\rm TC}$, where the widths of
$\;\eta'_T \rightarrow W_l^+ W_l^- \gamma,\, W_l^\mp Z_l^0 W_t^\pm,\,
   W_l^+ W_l^- Z_t^0 \;,\;$
with $l\,(t)$ referring to a longitudinal (transverse) polarisation,
have been combined as $\Gamma(\pi_T\pi'_T{\cal A})$  in the last column.
Here we have used the parameters $\;\alpha = e^2/(4\pi) = 1/137 \;,$
$\;s_W^2 = 0.232\;,$   $\;m_W = 80.2\,\rm GeV \;,\;$  and
$\;m_Z = 91.2\,\rm GeV\;.\;$
There are also decays into four longitudinal weak gauge bosons, which one might
naively expect to be significant because
the  equivalent decay amplitudes involving the $\pi_T^{\pm,0}$  contain the
$\rho_T$ in the intermediate states and no electroweak couplings.
It turns out that their widths are suppressed by the presence of the
antisymmetric tensor  $\epsilon_{\kappa\lambda\mu\nu}$  in the amplitudes and
by
the four-body final-state phase space.
For example, using (\ref{vpp}), (\ref{grpp}), (\ref{g4}), and (\ref{c2}),
with $\;N_{\rm TC} = 4\;$ we obtain
\begin{equation}
\Gamma_{\eta'_T \rightarrow W_l^+ W_l^- Z_l^0 Z_l^0} \sim 10^{-3}\,\rm MeV \;
\end{equation}
for the choice $\;2c_1-c_3/g^2=-2c_2/g \;,\;$ whose QCD analogue led to good
agreement with data for  $\;\omega\rightarrow 3\pi\;$  \cite{tf}.
(Since the actual values of $c_1$ and $c_3$ will not affect our results, we
leave them undetermined.)
We therefore expect that there are no other significant partial decay widths of
the $\eta'_T$   than those given in Table \ref{widths}.

Table \ref{events}  shows for   $\;N_{\rm TC} = 4,5,6\;$  the total width
$\Gamma_{\eta'_T}$ in each of the two  cases considered here.
In Case 1, where the $\eta'_T$ is coupled to fermion pairs, we obtain
$\Gamma_{\eta'_T}$
by summing all of the partial widths listed in Table \ref{widths}.
In Case 2, where the $\eta'_T$ has negligible couplings to fermion pairs,
$\Gamma_{\eta'_T}$ is  found from  Table \ref{widths}  by   combining only the
widths of the decays into electroweak  gauge bosons, and hence the $\eta'_T$
is a very narrow resonance.

We now turn to the production and detection of the $\eta'_T$ at a photon linear
collider.
By directing a low-energy laser beam at a high-energy $e^+(e^-)$ beam almost
head-to-head, a beam of backscattered photons is produced, carrying a large
fraction of the  $e^+(e^-)$-beam energy.
The resulting $\gamma\gamma$ beams have a luminosity comparable to that of the
parent $e^+e^-$ beams.
The energy-distribution function of a backscattered photon is given by
\cite{ifg}
\begin{equation}
f_{e/\gamma}(x) =
 {1\over D(\xi)}\Biggl[1-x+{1\over 1-x}-{4x\over\xi(1-x)} +
                      {4x^2\over \xi^2(1-x)^2}\Biggr]
\end{equation}
where
$$
D(\xi) = \Biggl(1-{4\over \xi}-{8\over \xi^2}\Biggl)\ln(1+\xi) +
         {1\over2} + {8\over \xi}-{1\over 2(1+\xi)^2} \;,
$$
$\xi = 4\omega_0 E_e/m_e^2 \;,\;$  with $\omega_0$ and $E_e$ being the energies
of the incident laser photon and the $e^+$ (or $e^-)$ beam, respectively, and
$\;x=\omega/E_e\;$ is the fraction of $E_e$ carried by the backscattered
photon.
$f_{e/\gamma}(x)$ vanishes for
$\;x>x_{\rm max} =\omega_{\rm max}/E_e= \xi/(1+\xi) \;.\;$
In order to avoid the creation of $e^+e^-$ pairs by the interaction of the
incident and backscattered photons, we require  $\omega_0 x_{\rm max}\le
m_e^2/E_e \;,\;$ which implies  $\;\xi \le 2+2\sqrt{2} \simeq 4.8 \;.\;$
For the choice $\;\xi = 4.8 \;,\;$ which maximizes $x_{\rm max}$, we obtain
$\;x_{\rm max}\simeq 0.83 \;,$   $\;D(\xi)\simeq 1.8 \;,\;$ and
$\;\omega_0 \simeq 0.31\,\rm eV\;$ for an $e^+e^-$ collider with center-of-mass
energy  $\;\sqrt{s_{e^+e^-}}=2\,\rm TeV \;.\;$
Here we have taken the photons to be unpolarised and the average number of
backscattered photons per positron (or electron) to be one.

Total cross sections $\sigma$ at the parent $e^+e^-$ collider are found by
folding  $\gamma\gamma$-subprocess cross sections
$\hat{\sigma}_{\gamma\gamma}$  with the photon distribution functions:
\begin{equation}
\sigma(s_{e^+e^-}) =
 \int_{\tau_1}^{\tau_2} d\tau
 \int_{\tau/x_{\rm max}}^{x_{\rm max}} {dx\over x}
  \;f_{e/ \gamma}(x) f_{e/ \gamma}(\tau/x) \;
\hat{\sigma}_{\gamma\gamma}(\tau s_{e^+e^-})
\end{equation}
where $\tau_1$ ($\tau_2$) is the minimum (maximum) of the range of
$\;\tau = s_{\gamma\gamma}/s_{e^+e^-}\;$ to be integrated over.
In order to find numbers of events, we  multiply $\sigma$ by  the yearly
integrated $e^+e^-$ luminosity $L_{ee}$.

In Case 1 the $\eta'_T$ decays almost entirely into a $t\bar{t}$ pair and so we
expect   $\;\gamma\gamma\rightarrow\eta'_T\rightarrow t\bar{t}\;$
to be the only channel likely  to be detectable.
The background for this channel comes from the process
$\;\gamma\gamma\rightarrow t\bar{t}\;.\;$
The combined amplitude for the signal and background processes is
\begin{eqnarray}
{\cal M}_{\gamma\gamma\rightarrow t\bar{t}}  &=&
-i{e^2 N_{\rm TC}\over 8\pi^2 F_{\pi_T}}\, {m_t\over F_{\pi_T}} \; \;
{ \epsilon^{\alpha\beta\mu\nu} \; k_\alpha k'_\beta
  \varepsilon_\mu (k) \varepsilon_\nu (k') \over
 s_{\gamma\gamma}-m_{\eta'_T}^2+im_{\eta'_T}\Gamma_{\eta'_T} } \;
\bar{u}(p) \gamma_5 v(p') \;
\nonumber \\  &&  +\;
\bigg( {2\over 3}e \bigg)^2  \, \bar{u}(p) \Bigg[
 \mbox{$\not{\!\varepsilon}$}(k)
  { 1\over \mbox{$\not{\!p}$} - \mbox{$\not{\!k}$} - m_t + i\epsilon }
   \mbox{$\not{\!\varepsilon}$}(k')
 +
 ( k \leftrightarrow  k' )
\Bigg] v(p')  \;,
\end{eqnarray}
where  $k,k'$  and  $p,p'$ are the four-momenta of the incoming photon pair and
outgoing $t\bar{t}$ pair, respectively, and
$\; s_{\gamma\gamma} = (k+k')^2 = (p+p')^2 \;.\;$
In the center-of-mass frame of the incoming photons the background $t\bar{t}$
production  is peaked in the forward and backward directions, whereas the
signal
$t\bar{t}$ are produced isotropically.
Hence we impose an angular cut $\;|\cos{\theta}| < 0.866\;$ where
$\theta$ is the scattering angle of the $t$'s in this frame.
We assume that the experimental resolution is smaller than  $\Gamma_{\eta'_T}$
and that the $t\bar{t}$ events can be fully reconstructed.
For $\;L_{ee}=10\,\rm fb^{-1} \;$ and $\;\sqrt{s_{e^+e^-}}=2\,\rm TeV \;,\;$
the number of  signal and background events in the mass interval
$\;m_{\eta'_T}-\Gamma_{\eta'_T}/2 < \sqrt{s_{\gamma\gamma}}
 < m_{\eta'_T}+\Gamma_{\eta'_T}/2\;$
is given in Table \ref{events}.\footnote{
The signal-background interference  contribution is small and has been included
in the number of background events.}
If desired, the background could be reduced further, while the signal being
increased, by employing polarized $\gamma\gamma$ beams as was done in Ref.
\cite{gh}  for Higgs production.

In Case 2 the channel
$\;\gamma\gamma\rightarrow\eta'_T\rightarrow\gamma\gamma\;$ is probably the
only
one likely to be viable.
The $WW$, $\gamma Z$, and $ZZ$ channels are known to have large backgrounds
\cite{gkps,gj}, while the $WW\gamma$ and $WWZ$ channels may be too small
to be useful.
The background in the $\gamma\gamma$ channel is dominated by $W$-boson loop
contributions.
The $\eta'_T$ being such a narrow resonance we may safely ignore the
interference effect between continuum background diagrams and the resonance
diagram.
Consequently, the  subprocess cross section can be written as a sum of a
resonance  cross  section and a continuum background cross section:
\begin{equation}
\hat{\sigma}_{\gamma\gamma\rightarrow\gamma\gamma }(s_{\gamma\gamma}) =
 \hat{\sigma}_{\gamma\gamma\rightarrow\gamma\gamma }^r(s_{\gamma\gamma}) +
 \hat{\sigma}_{\gamma\gamma\rightarrow\gamma\gamma}^b(s_{\gamma\gamma})  \;.
\end{equation}
The resonance cross section is given by
\begin{equation}
\hat{\sigma}_{\gamma\gamma\rightarrow\gamma\gamma}^r(s_{\gamma\gamma}) =
 8\pi\,
 {\Gamma_{\eta'_T \rightarrow \gamma\gamma}^2\over
  (s_{\gamma\gamma}-m_{\eta'_T}^2)^2 + m_{\eta'_T}^2 \Gamma_{\eta'_T}^2}  \;.
\end{equation}
The background cross section can be estimated by scaling the
$\;\gamma\gamma\rightarrow Z_t Z_t\;$ cross section calculated in
Ref.~\cite{gj}
by a factor $\;e^4/g_{WWZ}^4 = \sin^4\theta_W/\cos^4\theta_W \;.\;$
One finds
$\;\hat{\sigma}_{\gamma\gamma\rightarrow\gamma\gamma}^b(s_{\gamma\gamma})
\simeq 25\,\rm fb\;$ in the relevant  $\sqrt{s_{\gamma\gamma}}$ range.
For the same values of $L_{ee}$ and $\sqrt{s_{e^+e^-}}$ as before,
we show  in Table \ref{events} the number of signal and background events
in the  interval
$\;1.38\,{\rm TeV} < \sqrt{s_{\gamma\gamma}} < 1.40\,\rm TeV\;$
without  employing any cuts.

In conclusion, we have shown that in the two cases discussed above
the $\eta'_T$ can be observed above the backgrounds at a TeV $\gamma\gamma$
collider.
We learn from Tables \ref{widths} and \ref{events} that for larger $N_{\rm TC}$
the signal-to-background ratio is better because the $\eta'_T$ mass is smaller
and the branching ratio of the $\gamma\gamma$ decay mode is larger.
Although in this paper we have considered only two possible cases in the
simplest TC model,  similar  analyses can be made in more complicated models
in which the    $t\bar{t}$ or $\gamma\gamma$ decay mode is significant.
Hence the $\eta'_T$ has the potential to be a useful probe of its
subconstituents.
Our results above should give additional motivation for  developing a
backscattered-laser beam facility in the future.

\acknowledgments
The author thanks John F. Donoghue for many helpful discussions and
suggestions.
The author also thanks Barry R. Holstein for some information on the $\eta$.
This work was supported in part by the U. S. National Science Foundation.

\begin{table}
\caption{The mass and the partial decay widths of the $\eta'_T$ for different
values   of $N_{\rm TC}$.
In the fourth column $\Gamma(f\bar{f}$, no $t\bar{t})$ is the combined width of
the decays into all ordinary-fermion pairs excluding $t\bar{t}$.
In the last column $\Gamma(\pi_T\pi'_T{\cal A})$ is the combined width of
$\;\eta'_T \rightarrow W_l^+ W_l^- \gamma,\, W_l^\mp Z_l^0 W_t^\pm,\,
   W_l^+ W_l^- Z_t^0 \;.\;$ \label{widths}}
\begin{tabular}{cccccccccc}
$N_{\rm TC}$  &  $m_{\eta'_T}$  &  $\Gamma(t\bar{t})$  &
$\Gamma(f\bar{f}$, no $t\bar{t})$  &
$\Gamma(gg)$\tablenote{For the three different values of $m_{\eta'_T}$
listed here,  $\;\alpha_s (m_{\eta'_T}) = 0.084,\,0.086,\,0.088\;,\;$
respectively, obtained using the results of Ref.~\cite{alphas}.}  &
$\Gamma(WW)$  &  $\Gamma(\gamma\gamma)$  &  $\Gamma(\gamma
Z)$  &    $\Gamma(ZZ)$ &  $\Gamma(\pi_T\pi'_T{\cal A})$
\\
&  (TeV)  &  (GeV)  &  (MeV)  &  (MeV)  &  (MeV)  &  (MeV)  &  (MeV)  &  (MeV)
&  (MeV)
\\  \hline
4 & 1.39  &  80.4  &  58.3  &  56.3  &  19.3  &  4.8  &  3.8  &  3.5  &  8.1
\\
5 & 1.11  &  63.1  &  46.7  &  55.2  &  15.3  &  3.8  &  3.0  &  2.8  &  2.1
\\
6 & 0.927  &  51.3  &  38.9  &  54.2  &  12.5  &  3.2  &  2.5  &  2.3  &  0.7
\\
\end{tabular}
\end{table}

\begin{table}
\caption{
The total width of the $\eta'_T$ and the number of $t\bar{t}$ ($\gamma\gamma$)
events for the  signal $S$  and background $B$  in Case 1 (2) for different
values of $N_{\rm TC}$.
\label{events}}
\begin{tabular}{ccccccccc}
&&  \multicolumn{3}{c}{Case 1}  &&  \multicolumn{3}{c}{Case 2}
\\   \cline{3-5} \cline{7-9}
$N_{\rm TC}$ && $\Gamma_{\eta'_T}$ & $S$ & $B$ && $\Gamma_{\eta'_T}$ & $S$ &
$B$
\\
  &&  (GeV)  &&&& (MeV)  && $\;\;\;\;\;$
\\ \hline
4 &&  80.6  &  103  &  70  &&  39.5  &  28  &  3
\\
5 &&  63.2  &  147  &  97  &&  27.0  &  51  &  4
\\
6 &&  51.4  &  189  &  120  &&  21.2  &  70  &  4
\\
\end{tabular}
\end{table}


\begin{references}
\bibitem{js'} Our notation is somewhat similar to that used in Ref.~\cite{js}
for the analogous effective Lagrangian in QCD.
\end{references}
\end{document}